\documentclass[a4paper,11pt,english]{article}
\usepackage{amsfonts}
\usepackage{graphicx}
\usepackage{amsmath}
\usepackage{color}

\newcommand{\be}{\begin{equation}}
\newcommand{\ee}{\end{equation}}
\begin{document}


\begin{titlepage}
\begin{center}

\noindent{{\LARGE{String correlators in AdS$_3$ from FZZ duality}}}

\smallskip
\smallskip

\smallskip
\smallskip

\smallskip
\smallskip
\smallskip
\smallskip
\smallskip
\smallskip
\smallskip
\smallskip

\noindent{\large{Gaston Giribet}}

\smallskip
\smallskip

\smallskip
\smallskip

\smallskip
\smallskip

\centerline{Physics Department, University of Buenos Aires FCEyN-UBA and IFIBA-CONICET}
\centerline{{\it Ciudad Universitaria, pabell\'on 1, 1428, Buenos Aires, Argentina.}}

\end{center}

\bigskip

\bigskip

\bigskip

\bigskip

\begin{abstract}
Motivated by recent works in which the FZZ duality plays an important role, we revisit the computation of correlation functions in the sine-Liouville field theory. We present a direct computation of the three-point function, the simplest to the best of our knowledge, and give expressions for the $N$-point functions in terms of integrated Liouville theory correlators. This leads us to discuss the relation to the $H^+_3$ WZW-Liouville correspondence, especially in the case in which spectral flow is taken into account. We explain how these results can be used to study scattering amplitudes of winding string states in AdS$_3$. 
\end{abstract}
\end{titlepage}


\section{Introduction}

Originally formulated by Fateev, Zamolodchikov and Zamolodchikov as a conjecture in an unpublished work, the FZZ correspondence provided us with one of the most useful tools to investigate black holes in the stringy regime. This correspondence establishes the duality between the sine-Liouville field theory and the string worldsheet sigma model on the 2D black hole \cite{Jerusalem, Mandal}, the latter being described by the $SL(2,\mathbb{R})_k/U(1)$ WZW model \cite{Witten}; see also \cite{DVV, BB}. This is a special kind of strong/weak duality that can actually be thought of as a sort of T-duality \cite{Seiberg}, as it derives from the mirror symmetry between the $\mathcal{N}=2$ super-Liouville theory and the Kazama-Suzuki SCFT for $SL(2,\mathbb{R})_k/U(1)$ \cite{HK}.

As said, FZZ duality has shown to be a remarkably useful picture to study black holes in string theory. For example, it was what made possible to construct the KKK matrix model for the 2D black hole \cite{KKK}. It also provided a method to investigate the horizon structure in the stringy regime \cite{GIK}. FZZ duality was also used to study other backgrounds in string theory, besides black holes; among them, AdS$_3\times \mathcal{M}$ spacetimes \cite{GK}, cosmological models \cite{Nakayama} and other time-dependent scenarios \cite{HT}.

First taken as a widely accepted conjecture, the FZZ duality was finally proven by Hikida and Schomerus in \cite{HS} by resorting to the so-called $H_3^+$ WZW-Liouville correspondence \cite{Stoyanovsky, RT}, which had been first extended to higher genus \cite{HS2} and combined with the free field representation \cite{Giribet}. Previous arguments to prove the duality were given by Maldacena in \cite{Maldacena}, who proved that FZZ followed from the mirror symmetry \cite{HK} of the supersymmetric theories after a projection (GKO quotient of the R-symmetry) and decoupling of fermions (see Figure 1). 

One of the most interesting aspects of the FZZ duality is that it relates models with different topologies. While the target space interpretation of the sine-Liouville CFT has topology $\mathbb{R}\times S^1$, the Euclidean 2D black hole corresponds to the so-called cigar geometry, with topology $\mathbb{R}^2$ (see Figure 2 below). This permits to investigate the common features shared by stringy backgrounds with quite diverse geometrical interpretations, and to related physical mechanisms that, a priori, are seemingly disconnected. One example is the mechanisms by means of which the winding number is violated in scattering processes in the black hole geometry and in the sine-Liouville theory. While in the former this is explained in terms of the topology of the cigar geometry, in the latter the violation is explained by the sine-Liouville potential and so it is dynamical. 
\begin{figure}[h]
    \begin{center}
        \includegraphics[width=15cm]{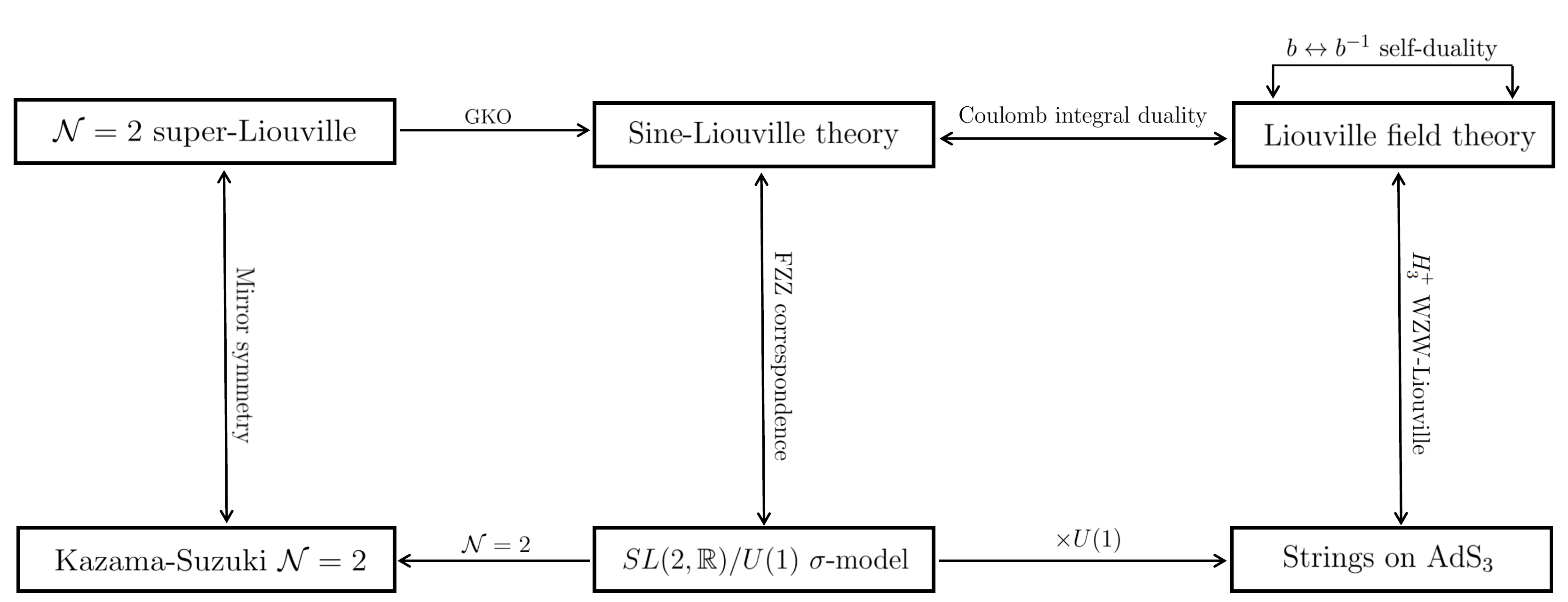}
        \caption{Scheme of different (duality) relations among CFTs. The FZZ correspondence relates sine-Liouville theory with the $SL(2,\mathbb{R})/U(1)$ coset $\sigma $-model that describe string theory on the Euclidean 2D black hole. The latter model can be used to describe string theory on AdS$_3$, since the spectral flow sectors of the $SL(2,\mathbb{R})$ WZW theory are conveniently realized in terms of the product $SL(2,\mathbb{R})/U(1)\times U(1)$. The latter theory also relates to the standard Liouville theory by the $H_3^+$ WZW-Liouville correspondence. Finally, the sine-Liouville correlators can be written as integrated Liouville correlators due to remarkaby identities among conformal integrals.}
        \label{Ricci}
    \end{center}
\end{figure}

In \cite{GIK2}, the FZZ duality was generalized, and it was shown there that the connection between states with non-trivial winding number and momentum in both theories is much more general than what it was originally considered. This can be seen not only at the level of the spectrum, but also at the level of interactions \cite{Giribet:2016ouf}. More recently, the FZZ duality has been considered in new contexts: In \cite{Jafferis}, for example, Jafferis and Schneider used it in their study of the ER=EPR correspondence. FZZ also appears mentioned in the analysis of black holes near the Hagedorn temperature that Maldacena and Chen perform in \cite{Chen}. Here, motivated by this revival of the FZZ duality, we revisit the computation of three-point correlation functions in the sine-Liouville theory. These observables have been discussed long time ago by Fukuda and Hosomichi \cite{FH}, who did a very interesting analysis of the residues of such three-point functions by proposing a specific contour prescription. We will explain the connection between our result and that of \cite{FH}. Besides, we will study the four-point correlation function and give an expression of it in terms of Liouville field theory correlators. This will lead us to discuss the relation to the generalized $H_3^+$ WZW-Liouville correspondence \cite{RT, HS2, Ribault}, which connects $SL(2,\mathbb{R})_k$ WZW correlation functions with integrated higher-point functions in standard Liouville theory. Then, we will explain how these results can be used to study scattering amplitudes in AdS$_3$ that involve string winding states.

\section{Sine-Liouville and FZZ}

\subsubsection{The conformal field theory}

The sine-Liouville field theory is defined by the action
\begin{equation}
S_{\text{sine-L}}[\lambda ]=\frac{1}{2\pi }\int d^{2}z\left( \partial \phi \bar{%
\partial }\phi +\partial X\bar{\partial }X+\frac{1}{\sqrt{2}}{R \,{b} \phi }+4\pi \lambda \ 
e^{\frac{1}{\sqrt{2}{b}}\phi }\cos ( \sqrt{{k}/{2}}\tilde{X})\, \right)  . 
\label{sL}
\end{equation}
This can be interpreted as the sine-Gordon model coupled to Liouville field theory, and hence the name. Field $\phi (z,\bar{z}) $ is a Liouville-like scalar field that takes values on $\mathbb{R}$ and has a background charge ${b} = -1/\sqrt{k-2}$. The field $X (z,\bar{z})$ lives on a circle of radius $\sqrt{k}$. This means that the target interpretation of the model is that of an Euclidean space with cylindrical topology. The field $\tilde X (z,\bar z )= X_L (z) - X_R (\bar{z})$ that appears in the interaction term is the T-dual of $X (z,\bar z ) = X_L (z) + X_R (\bar z )  $.

Action (\ref{sL}) defines a 2-dimensional non-compact CFT with central charge 
\begin{equation}
c=2+\frac{6}{k-2} \label{Yc}\, .
\end{equation}
It admits the interpretation of a string theory $\sigma$-model on a 2-dimensional linear dilaton background in the presence of a non-homogeneous tachyon condensate, the tachyon potential being given by the interaction term. The latter resembles a Liouville wall that prevents the strings from exploring the strong coupling region, with the effective string coupling being $g_s(\phi )=\exp (-\sqrt{2}b\phi )$. In fact, due to the presence of the linear dilaton term, the zero mode of $\phi $ comes along with the Euler characteristic of the worldsheet surface, so producing the correct exponent in the genus expansion. The positive constant $\lambda $ is associated to the value of the effective string coupling at certain position $\phi $, as its value can be set to $\lambda =1$ by simply rescaling the zero mode of $\phi $.

\subsubsection{FZZ duality}

FZZ duality states that the model defined by action (\ref{sL}) is dual to string theory on the Euclidean 2D black hole background. The latter is also called the cigar geometry, and it is described by the $SL(2,\mathbb{R})_k/U(1)$ gauged WZW model \cite{Witten}. This duality is surprising for many reasons: Firstly, as we said, it relates two models with different topology. While sine-Liouville is defined on a space with topology $\mathbb{R}\times S^1$, the Euclidean black hole resembles a semi-infinite cigar and so it has topology $\mathbb{R}^2$. Secondly, the two models also differ in the mechanism by means of which the strings are prevented from exploring the strong coupling region. Both models are asymptotically $\mathbb{R}\times S^1$, and, for both, the string coupling constant vanishes there. However, while in sine-Liouville theory what prevents the strings from entering into the strong coupling region is a potential wall, in the case of the cigar geometry such region is inaccessible simply because it lies behind the horizon, a region that is not part of the Euclidean manifold. In the case of the cigar, the asymptotically $\mathbb{R}\times S^1$ region is far from the horizon --which corresponds to the tip of the cigar--, and the $S^1$ direction corresponds to the Euclidean time. A third difference between the sine-Liouville theory and the Euclidean black hole theory is the mechanism producing the violation of the winding number in the string interaction processes. While for the former this is explained by the presence of an explicit dependence of the dual $\tilde X$ field in the Lagrangian, in the latter this is merely due to topology. 

FZZ has been reviewed in various works; see for instance \cite{KKK, GIK2, Giveon2} and references therein and thereof. 
\begin{figure}[h]
    \begin{center}
        \includegraphics[width=9cm]{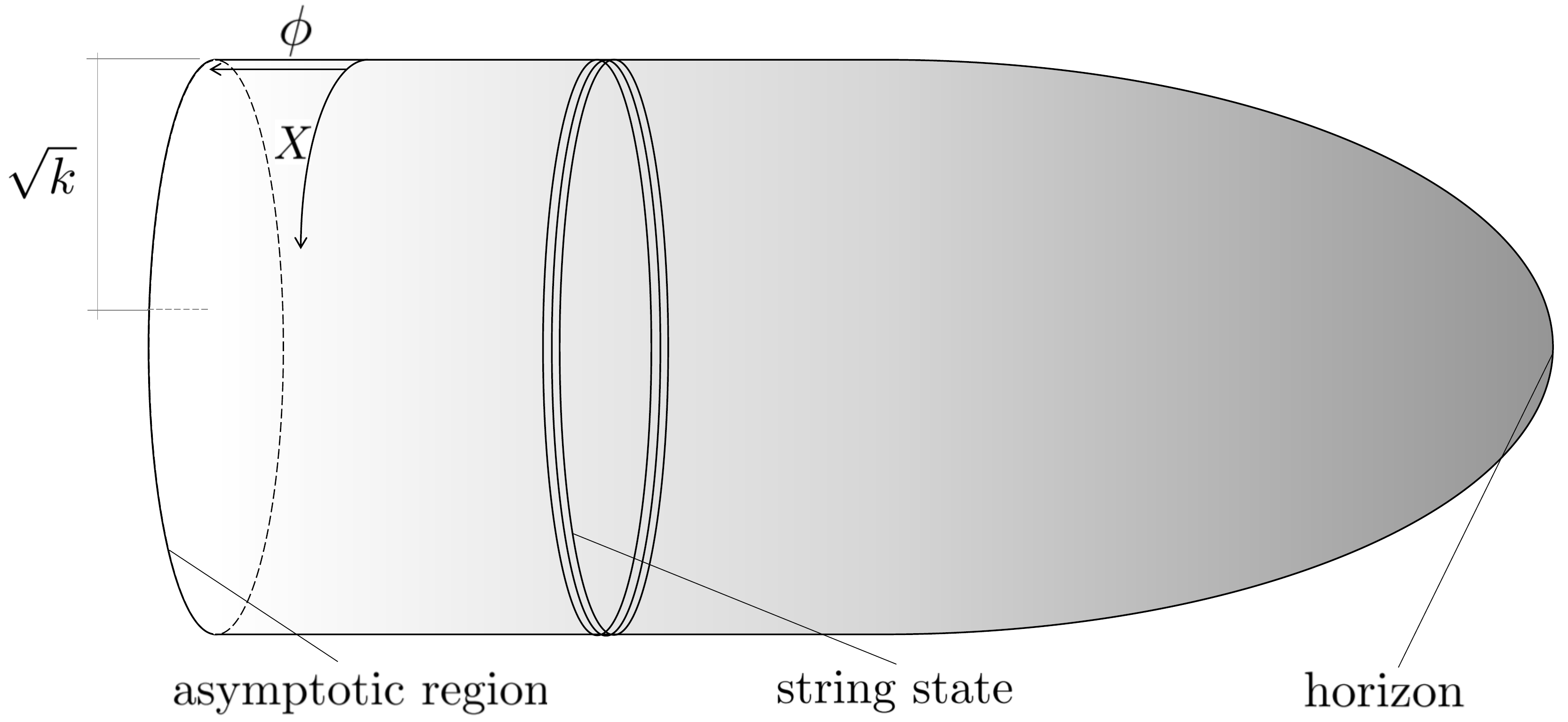}
        \caption{The geometry of the Euclidean 2D black hole resembles a semi-infinite cigar of radius $\sqrt{k}$ in string units ($\alpha' = 1$ here). The black hole horizon corresponds to the tip of the cigar. The string non-linear $\sigma $-model on this space is given by the $SL(2,\mathbb{R})_k/U(1)$ WZW model. The theory on AdS$_3$ can be described by tensoring $SL(2,\mathbb{R})_k/U(1)\times {\text{time}}$.}
        \label{Ricci}
    \end{center}
\end{figure}

\subsubsection{Spectrum}

The primary states of sine-Liouville CFT are created by exponential vertex operators of the form
\begin{eqnarray}
V^{\ p,\omega }_{j}(z) = e^{\sqrt{\frac{2}{k-2}}j\phi(z,\bar{z})} e^{i\sqrt{\frac{2}{k}}p X(z,\bar{z})+i\sqrt{{2k}}\omega \tilde{X}(z,\bar{z})} \ , \label{Y23}
\end{eqnarray}
where $j, p$, and $\omega $ are quantum numbers labeling the momenta along the direction $\phi $, the momentum along the compact direction $X$, and the winding number around the latter, respectively. While $p$ is conserved, the winding number $\omega $ is not; the latter can be associated to a dual momentum in $\tilde X$; cf. \cite{GN3, MO3, FH, Yo2, Yo}.

In order to make the connection with the Euclidean black hole spectrum clear, it is convenient to define the variables 
\begin{equation}
m = \frac{1}{2} (p + k\omega ) \ , \ \ \ \ \ \ \bar{m} = \frac{1}{2} (p - k\omega ) \ . \label{Y24}
\end{equation}
In the $SL(2,\mathbb{R})_k/U(1)$ WZW model, which describes the 2D black hole $\sigma$-model, the variables $j$, $m$ and $\bar m$ label the unitary representations of $ SL(2,\mathbb{R})$ starting from which the whole perturbative string spectrum is constructed. $k$ is the Kac-Moody level of the WZW model. In this coset model, the winding number, which counts how many time a string state winds around the asymptotic cylinder, is given by $\omega = -(m+\bar m)/k $, while the momentum around the compact direction is $p=m-\bar m$; recall that the relation to sine-Liouville is a T-dual duality. Winding number will in general not be conserved. As said, this is due to the topology of the geometry.  

The conformal dimensions of the primary states created by (\ref{Y23}) are given by
\begin{equation}
h = -\frac{j(j+1)}{k-2} + \frac{m^2}{k} \ , \ \ \ \bar h = -\frac{j(j+1)}{k-2} + \frac{\bar{m}^2}{k} \ . \label{Y26}
\end{equation}
Notice that these expressions are invariant under the changes $j\to -1-j$, $m \to -m$ and $\bar m \to - \bar m$. In terms of the variables (\ref{Y24}), the vertex operators (\ref{Y23}) take the form
\begin{eqnarray}
V_{j,m,\bar{m}}(z) = e^{\sqrt{\frac{2}{k-2}}j\phi(z,\bar{z})} e^{i\sqrt{\frac{2}{k}}mX(z)+i\sqrt{\frac{2}{k}}\bar{m}\bar{X}(\bar{z})}\label{Y25}
\end{eqnarray}
where, for short, we denote $X(z)=X_L(z)$ and $\bar{X}(\bar{z})=X_R(\bar{z})$. 

The interaction term in (\ref{sL}) can be written as the sum of two particular vertices (\ref{Y25}), with the values of the momenta corresponding to marginal operators ($h=\bar h = 1$); namely
\begin{equation}
{2\lambda } \int d^2z \ \cos (\sqrt{{k}/{2}} \, \tilde X (z)\, ) = {\lambda } \int d^2z  \ V_{1-\frac{k}{2},\frac{k}{2},-\frac{k}{2}} (z) + {\lambda } \int d^2z \ V_{1-\frac{k}{2},-\frac{k}{2},\frac{k}{2}} (z) . \label{Y27}
\end{equation}
Correlation functions can be computed by inserting these interaction operators as screening charges in the Coulomb gas approach.

\subsubsection{Correlation functions}

Correlation functions in sine-Liouville model are defined as follows
\begin{equation}
\mathcal{A}^N _{\text{sine-L}}\equiv  \Big\langle  \prod_{i=1}^{N} \ V_{j_i , m_i, \bar{m}_i} (z_i) \ \Big\rangle_{\text{sine-L}} = \int {\mathcal D}\phi \, {\mathcal D}X \, e^{-S_{\text{sine-L}}[\lambda ]}\ \prod_{i=1}^{N} V_{j_i , m_i, \bar{m}_i} (z_i) 
\end{equation}
which, by means of standard techniques \cite{diFK, GL, BB}, can be written as
\begin{eqnarray}
\mathcal{A}^N_{\text{sine-L}} &=& {\lambda^{s}}\, \Gamma (-s)
\int_{\mathbb{C}^{s}} \prod_{l=1}^{s} d^2w_l \int {\mathcal D}\phi {\mathcal D}X  e^{-S_{\text{sine-L}}[\lambda =0]}
\, \Big[ \, \prod_{i=1}^N V_{j_i , m_i, \bar{m}_i} (z_i) 
 \nonumber \\
&& \ \ \ \ \prod_{l=1}^{s} \Big(    V_{1-\frac{k}{2},\frac{k}{2},-\frac{k}{2}} (w_l) + V_{1-\frac{k}{2},-\frac{k}{2},\frac{k}{2}} (w_l)  \Big) \, \Big] \label{Y210bis}
\end{eqnarray}
with the following condition
\begin{eqnarray}
\sum_{i=1}^{N}j_i + 1 - \frac{k-2}{2} \, s = 0 \ . \label{JK}
\end{eqnarray}

Formula (\ref{Y210bis}) shows that the computation of sine-Liouville $N$-point correlation functions reduces to that of $(N+s)$-point correlation functions of a free theory ($\lambda =0$) with a background charge. The latter can in principle be computed resorting to the Coulomb gas formalism. However, before doing so, let us explain the prefactors appearing in the formula above: The factor $\lambda^{s}$ comes from the expansion of the exponential of the interaction term, along with the insertion of the $m+n$ operators that come to screen the background charge. Notice that (\ref{JK}) implies that, in the case of the spherical partition function ($N=0$), for the value of the Kac-Moody level $k=9/4$, which is the value that renders the theory critical ($c=26$), one obtains $Z\sim \lambda ^8$, in perfect agreement with the Knizhnik-Zamolodchikov-Polyakov (KPZ) scaling discussed in \cite{KKK}; see Eq. (5.10) therein. The factor $\Gamma (-s)$ in (\ref{Y210bis}) comes from the integration over the zero mode of $\phi $. The interpretation of this factor and the divergences it produces when $s\in \mathbb{Z}_{\geq 0}$ is similar to the one given in \cite{diFK} for resonant correlators in Liouville theory; see Eq. (2.10)-(2.12) therein, cf. Eq. (3.3) of \cite{GL}.

Expression (\ref{Y210bis}) includes the possibility of having correlators that do not conserve the total winding number $\Delta \omega \equiv \sum_{i=1}^N\omega_i$. This is realized by splitting the interaction operator in $s=m+n$, with $m$ operators $V_{1-\frac k2, \frac k2, -\frac k2}$ and $n$ operators $V_{1-\frac k2, -\frac k2, \frac k2}$. Then, we have
\begin{eqnarray}
\sum_{i=1}^{N}m_i + \frac{k}{2} (m-n)= 0 \ , \ \ \ \ \sum_{i=1}^{N}\bar{m}_i + \frac{k}{2} (n-m)= 0 \, ;
\end{eqnarray}
that is to say,
\begin{equation}
 \sum_{i=1}^N p_i = 0 \ , \ \ \ \ \sum_{i=1}^N \omega_i = n-m \, .\label{montoya}
\end{equation}
In this case, the integration over the zero mode of $\phi$ also selects the number of screening charges $m$, $n$ that will contribute to non-vanishing correlators. We see from (\ref{montoya}) that the violation of the total winding number $\Delta \omega $ is controlled by the difference $n-m$. Therefore, we can compute correlators that represent processes with a specific total winding number; namely 
\begin{eqnarray}
\mathcal{A}^{N,\Delta \omega }_{\text{sine-L}} =  {\lambda^{m + n}} &&\frac{\Gamma (-m-n)\Gamma(m+n+1)}{\Gamma(n+1)\Gamma(m+1)} 
\int_{\mathbb{C}^{m}} \prod_{l=1}^{m} d^2u_l \int_{\mathbb{C}^{n}} \prod_{r=1}^{n} d^2v_r 
\int {\mathcal D}\phi \, {\mathcal D}X \,  e^{-S_{\text{sine-L}}[\lambda =0]}
 \nonumber \\
&&\ \ \prod_{i=1}^N V_{j_i , m_i, \bar{m}_i} (z_i) \prod_{l=1}^{m} V_{1-\frac{k}{2},\frac{k}{2},-\frac{k}{2}} (u_l)
\prod_{r=1}^{n} V_{1-\frac{k}{2},-\frac{k}{2},\frac{k}{2}} (v_r), \label{Y210}
\end{eqnarray}
where the new superscript $\Delta \omega$ indicates the total winding number. The factor $\frac{\Gamma(m+n+1)}{\Gamma(n+1)\Gamma(m+1)}$ comes from the combinatorics $\big(\begin{smallmatrix}
  m+n \\
  n 
\end{smallmatrix}\big)$, counting all combinations of $m,n$ screening operators $V_{1-\frac k2 , \pm \frac k2 , \mp \frac k2}$ that contribute to a process with a definite $\Delta \omega $ in the Cauchy product when going from (\ref{Y210bis}) to (\ref{Y210}).

The next step is to compute the OPE among the $N+m+n$ operators in the free theory ($\lambda=0$) to work out the Wick contractions in (\ref{Y210}). Considering the OPE
\begin{equation}
V_{j_i,m_i,\bar{m}_i} (z_i) V_{j_j,m_j,\bar{m}_j} (z_j) \simeq (z_i - z_j)^{-\frac{2}{k-2}j_ij_j+\frac{2}{k}m_im_j} (\bar{z}_i - \bar{z}_j)^{-\frac{2}{k-2}j_ij_j+\frac{2}{k}\bar{m}_i\bar{m}_j}
\end{equation}
we easily find
\begin{eqnarray}
\mathcal{A}^{N,\Delta \omega}_{\text{sine-L}} &=& {\lambda^{m+n}}\,  \frac{\Gamma (-m-n)\Gamma(m+n+1)}{\Gamma(n+1)\Gamma(m+1)}
\prod_{i<j}^N (z_i - z_j)^{\frac{2}{k}m_im_j-\frac{2}{k-2}j_ij_j} (\bar{z}_i - \bar{z}_j)^{\frac{2}{k}\bar{m}_i\bar{m}_j-\frac{2}{k-2}j_ij_j}  \nonumber \\
&& \int_{\mathbb{C}^{m}} \prod_{l=1}^{m} d^2u_l \int_{\mathbb{C}^{n}} \prod_{r=1}^{n} d^2v_r \,
\prod_{l=1}^{m}\prod_{r=1}^{n} |u_ l - v_r|^{2-2k}
\prod_{l<l'}^{m} |u_l - u_{l'}|^{2}
\prod_{r<r'}^{n} |v_r - v_{r'}|^{2}
\nonumber \\
&&\prod_{i=1}^N \prod_{l=1}^{m} (z_i - u_l)^{j_i+m_i} (\bar{z}_i - \bar{u}_l)^{j_i-\bar{m}_i}
\prod_{i=1}^N \prod_{r=1}^{n} (z_i - v_r)^{j_i-m_i} (\bar{z}_i - \bar{v}_r)^{j_i+\bar{m}_i} 
. \label{Integral}
\end{eqnarray}
This is the integral representation of the genus-zero sine-Liouville correlation functions. Below, we will review how the tree-level scattering amplitudes in AdS$_3$ can be obtained from these observables. 

\subsubsection{Strings theory on AdS$_3$}

There exists a close relation between the worldsheet string theory on the 2D black hole and that of the theory on AdS$_3$ space(time). The non-linear $\sigma$-model that defines string theory on Euclidean AdS$_3$ with pure NS-NS fluxes is given by the gauged WZW model on the homogeneous space $H^+_3=SL(2,\mathbb{C})/SU(2)$, with the WZW level being $k=R^2/\alpha '$, with $R$ the radius of AdS$_3$. The theory on the Lorenztian AdS$_3$ corresponds to the WZW model on $SL(2,\mathbb{R})$, which is supposed to be obtained from the Euclidean case via analytic continuation. This theory has been extensively studied in the past; see \cite{MO1, MO2, MO3} and references therein. Recently, this model received renewed attention, and there have been many interesting developments; see for instance \cite{Alfa}-\cite{Omega}; see also \cite{DE1, DE2} for an interesting recent study of the AdS$_3$ string correlators.

The spectrum of the Lorentzian theory is given in terms of a subset of unitary representations of $SL(2,\mathbb{R})$ and their Kac-Moody affine extensions \cite{MO1}. These representations are built up starting from the discrete and continuous series of $SL(2,\mathbb{R})_k\otimes SL(2,\mathbb{R})_k$, labeled by indices $j, m$ and $j,\bar m$. An additional number $\omega \in \mathbb{Z}$ comes to label the spectral flow sectors and it physically represents the string winding number. The values $j\in -\frac 12 +i\mathbb{R}$ correspond to the continuous series representations of $SL(2,\mathbb{R})$ and describe the so-called long string states, which have continuous energy spectrum. On the other hand, the values $-\frac 12 < j <\frac{k-3}{2}$ correspond to a subset of the highest and lowest weight representations and describe the short strings, with discrete spectrum; see \cite{MO1} for details (notice that, when comparing with the conventions in \cite{MO1}, it is necessary to make $j\to j+1$). 

A convenient way of describing the theory on AdS$_3$ is to consider the product $SL(2,\mathbb{R})_k/U(1) \, \times \, U(1)$, where the first factor corresponds to the Euclidean black hole, which contributes with (\ref{Y26}), and the extra $U(1)$ is a timelike scalar with momentum $m+\frac k2 \omega$, cf. \cite{MO1,GN2}. This gives the central charge of the worldsheet theory on AdS$_3$, i.e. $c=3k/(k-2)$, and it also gives the correct worldsheet conformal dimensions of the string states in AdS$_3$; namely
\begin{equation}
h = -\frac{j(j+1)}{k-2} - m\omega -\frac k4 \omega ^2 \ , \ \ \ \bar h = -\frac{j(j+1)}{k-2}  - \bar m \omega -\frac k4 \omega ^2 \ . \label{Y26ghjhgf}
\end{equation}
The imaginary part of the label $j$ is associated to the radial momentum in AdS$_3$ -- in $\sqrt{k}$ units--, while $m-\bar m$ and $m+m+k\omega$ are the angular momentum and the energy, respectively. In the case of long strings, the spectral flow number $\omega $ is the string winding number around the boundary of AdS$_3$; and, in contrast to the coset theory, in the theory on AdS$_3$ the quantum number $\omega $ is independent of $m$ and $\bar m$. In general, the winding number will not be conserved, as the target space is simply connected. Indeed, there are scattering processes that produce the change of the total winding number, cf. \cite{GN3, MO3, Yo, Yo2}. 

The $N$-point amplitudes of winding string states in AdS$_3$ can be obtained from those of the coset theory by multiplying the correlators associated to the latter by a factor 
\begin{equation}
\lim_{z_N\to \infty }\, |z_N|^{4}\, \prod_{1\leq l <l'}^{N}|z_l - z_{l'}|^{\frac{4}{k} (m_l+\frac k2 \omega_l)(m_{l'}+\frac k2 \omega_{l'})}, \label{coconor17}
\end{equation}
and then integrating over $N-3$ worldsheet insertions, having previously fixed $z_1=0$, $z_2=1$ and $z_N\to\infty $ to cancel the volume of the conformal Killing group, $PSL(2,\mathbb{C})$. Factor (\ref{coconor17}) is the contribution of the extra timelike $U(1)$. Therefore, via FZZ duality, we can write string scattering amplitudes of winding strings on Lorentzian AdS$_3$ in terms of correlation functions in sine-Liouville field theory by simply including (\ref{coconor17}). The string coupling constant $g_s$ in AdS$_3$ is given in terms of $\lambda $, cf. \cite{GK}.

\section{The 3-point function}

\subsubsection{Coulomb gas representation}

Let us consider first the 3-point function ($N=3$) in sine-Liouville theory with a given winding number $\Delta \omega $. According to what we discussed above, this can be written as
\begin{eqnarray}
\mathcal{A}_{\text{sine-L}}^{N=3,\Delta \omega } &=& \lambda^{n+m}\frac{\Gamma (-n-m) \Gamma (1+n+m)}{\Gamma (n+1)\Gamma(m+1)}\int_{\mathbb{C}^{m}} \prod_{a=1}^{m}d^2u_a \int_{\mathbb{C}^{n}} \prod_{i=1}^{n}d^2v_i \, \prod_{a=1}^{m} \Big[ (u_a)^{j_1+m_1}(\bar u_a)^{j_1-\bar m_1}\nonumber \\
&&\ \ (1-u_a)^{j_2+m_2}(1-\bar u_a)^{j_2-\bar m_2}
\Big]\,
\prod_{i=1}^{n} \Big[ (v_i)^{j_1-m_1}(\bar v_i)^{j_1+\bar m_1}
(1-v_i)^{j_2-m_2}(1-\bar v_i)^{j_2+\bar m_2}
\Big]\,
\nonumber \\
&& \ \ \prod_{1\leq a<a'}^{m}  |u_a-u_{a'}|^{2} \, \prod_{1\leq i<i'}^{n}  |v_i-v_{i'}|^{2} \,
\, \prod_{a=1}^{m} \prod_{i=1}^{n}  |u_a-v_{i}|^{2-2k} \, ,\label{DFGHJK}
\end{eqnarray}
where we are using projective invariance to set $z_1=0$, $z_2=1$, and $z_3=\infty $. Recall that non-vanishing correlators must satisfy
\begin{equation}
    \sum_{i=1}^{3}j_i+1=\frac{k-2}{2}\, (m+n)
\end{equation}
together with
\begin{equation}
    \sum_{i=1}^{3}p_i= \sum_{i=1}^{3}(m_i+\bar m_i)=0\ ,  \ \ \ \ \ \Delta \omega \equiv \sum_{i=1}^{3}\omega _i= \frac 1k\, \sum_{i=1}^{3}(m_i-\bar m_i)=(n-m) \, .
\end{equation}

Here, we are concerned with correlators that involve states with non-vanishing winding numbers. A particularly interesting case is $ \Delta \omega =1$, which corresponds to $n=m+1$. This would give the result for the 3-point scattering amplitude of a process in AdS$_3$ that violates the winding number. This observable has been computed in the literature by different methods. In \cite{GN3}, a free field conjugate representation for the theory on $SL(2,\mathbb{R})_k/U(1)\times U(1)$ was proposed and used to compute correlators in the Coulomb gas formalism; in \cite{MO3}, an auxiliary operator with winding number 1 -- the so-called spectral flow operator-- was introduced in order to produce the violation of the total winding number; in \cite{Yo2}, discrete symmetries of the set of solutions of the Knihznik-Zamolodchikov equations were used to obtain the same result. Here, we will follow a rather different approach: We will compute the 3-point function with $\Delta \omega =1$ directly in the sine-Liouville theory and infer from FZZ the result for AdS$_3$. This will permit us to make contact with other aspects of WZW correlators; for example, we will see that our computation in the sine-Liouville theory turns out to be in agreement with the generalization of the $H_3^+$ WZW-Liouville correspondence worked out in \cite{Ribault}, where spectral flow was included in the scheme. This might seem to be expected since, after all, the FZZ correspondence has been proven in \cite{HS} precisely using the $H_3^+$ WZW-Liouville correspondence. However, there is a caveat here: One has to be reminded of the fact that here, in contrast to \cite{HS}, we are considering correlators with non-vanishing winding numbers.

The residues of the 3-point function of winding states in sine-Liouville theory had already been computed in \cite{FH}. Here, we will complement that computation by carefully keeping track of the factors that come from the integration over the zero mode of the Liouville type field. This makes a difference w.r.t. \cite{FH} as it changes the pole structure of the final result. Our result turns out to be in agreement with the results of \cite{GN3, MO3, Ribault}. Another difference w.r.t. the computation in \cite{FH} is that we will avoid dealing with the contour prescription used therein to solve the multiple integral (\ref{DFGHJK}). Instead, we will show how the integrals involved can be converted into a standard Dotsenko-Fateev integral as those studied in the Minimal Models. This makes our computation independent of specific contour prescriptions.

\subsubsection{Dotsenko-Fateev integral and DOZZ formula}

Before showing that the integral (\ref{DFGHJK}) can be transformed into a Dotsenko-Fateev integral, let us recall the definition of the latter together with some of its most salient properties. Dotsenko-Fateev integral is a multiple conformal integral of the Selberg type. More specifically, it can be regarded as a generalization of the Shapiro-Virasoro amplitude. It is defined as
\begin{eqnarray}
I (\alpha , \beta ; m) = \int_{\mathbb{C}^{m}} \prod_{a=1}^{m} d^2u_a \, \prod_{a=1}^{m} \, \Big[ \, |u_a|^{2\alpha} |1-u_a|^{2\beta}\, \Big] \, 
\prod_{1\leq a<a'}^{m}|u_a-u_{a'}|^{4\sigma} .\label{aboje}
\end{eqnarray}
Remarkably, this integral can be exactly computed \cite{DF} and shown to admit a relatively simple form: It can be expressed in terms of quotients and products of $\Gamma$-functions. Besides, as we will review below, it can also be expressed in terms of the special function $\Upsilon_b$, which is usually employed in Liouville field theory (see (\ref{HUPA}) below). In fact, integral (\ref{aboje}) appears in the Coulomb gas realization of Liouville field theory \cite{GL}, which, after a proper analytic continuation in $m$, yields a result in agreement with the DOZZ formula for the Liouville structure constants \cite{ZZ, DO}. More precisely, if we denote the Liouville structure constant $C(\alpha_1 , \alpha_2, \alpha_3)$, then we have
\begin{eqnarray}
   \Big\langle   \ \prod_{i=1}^3 \, V_{\alpha_i} (z_i) \ \Big\rangle_{\text{L}} \, =\, \prod_{1\leq i<i'}^3\,|z_i-z_{i'}|^{2\sum_{l=1}^3\Delta_l-4(\Delta_i + \Delta_{i'})}\, \, C(\alpha_1 , \alpha_2, \alpha_3)  
\end{eqnarray}
with
\begin{eqnarray}
C(\alpha_1 , \alpha_2, \alpha_3) \, = \, b\, \Gamma (-m) \,{\tilde \mu}^m \, I (-\alpha_1/b , -\alpha_2/b ; m) \, ,\label{dis}
\end{eqnarray}
where $m=b^2+1-b\sum_{i=1}^3\alpha_i$, and where we are using the standard notation to describe the $c>25$ Liouville field theory (cf. \cite{ZZ}); namely, we consider the exponential primary operators $V_{\alpha_i}(z_i)=\exp(\sqrt{2}\alpha_i \varphi (z_i))$ of conformal dimension $\Delta_i= \alpha_i(Q-\alpha_i)$, with the Liouville central charge $c=1+6Q^2$, $Q=b+1/b$. Again, we can consider $z_1=0$, $z_2=1$, $z_3=\infty $. The constant $\tilde \mu$ is the so-called dual Liouville cosmological constant, and it relates to the usual cosmological constant, $\mu$, by the equation $\tilde \mu = \frac{1}{\pi}\gamma (1-b^{-2}) (\pi\mu \gamma(b^2))^{1/b^2}$, with $\gamma(x)\equiv \Gamma(x)/\Gamma(1-x)$.

Before proving that (\ref{dis}) reproduces the DOZZ formula \cite{ZZ,DO}, we have to specify how to integrate in (\ref{aboje}). That is a set of integrals over the whole complex plane $\mathbb{C}$. The measure is $ d^2u_a = \frac i2\, du_a d\bar{u}_a$, which we can separate in real and imaginary parts, $u_a = x_a + i y_a$, $\bar{u}_a = {x}_a - i {y}_a$. In order to integrate, first it is convenient to Wick rotate $x_a \to ix_a$ and then introduce a deformation parameter $\varepsilon $ by defining $| u_a |^2= -x_a ^2+y_a ^2+i\varepsilon $; this permits to avoid the poles at $x_a =\pm y_a $. Finally, we can define coordinates $x^{\pm }_a = \pm  x_a + y_a $ and integrate over $x^-_a$ while keeping the $x^+_a$ fixed. Iterating this procedure and taking care of the combinatorics, one arrives to the formula 
\begin{eqnarray}
 &&\int_{\mathbb{C}^{m}} \prod_{a=1}^{m} d^2u_a \, \prod_{a=1}^{m} \Big( |u_a|^{2\alpha} |1-u_a|^{2\beta} \Big) \, 
\prod_{1\leq a<a'}^{m}|u_a-u_{a'}|^{4\sigma} =  
\Gamma(m+1)\, \pi^m \, (\gamma (1-\sigma))^m \nonumber \\
&&\ \ \prod_{r=1}^{m}\Big[ \gamma (r\sigma )
\gamma (1+\alpha +(r-1)\sigma )
\gamma (1+\beta +(r-1)\sigma )
\gamma (-1-\alpha -\beta -(m-2+r)\sigma )
\Big]\nonumber\, ;
\end{eqnarray}
see Eq. (B.9) of \cite{DF}.

As said, the formula above can be regarded as a multiple generalization of the Shapiro-Virasoro type integral formula: For $m=1$ this reduces to the well-known result
\begin{eqnarray}
\int_{\mathbb{C}} d^2u \, \, |u|^{2\alpha}\, |1-u|^{2\beta}\, = \, \pi \, \frac{\Gamma (1+\alpha )\Gamma (1+\beta )\Gamma (1-\alpha- \beta )}{\Gamma(-\alpha)\Gamma(-\beta)\Gamma (\alpha + \beta )}\, .
\end{eqnarray}

Now, we are ready to review how to obtain the DOZZ formula from the integral above. More precisely, what we want to prove is that the following identity holds
\begin{eqnarray}
&&b\, \Gamma(-m)\, \int_{\mathbb{C}^{m}} \prod_{a=1}^{m} d^2u_a \, \prod_{a=1}^{m} \Big( |u_a|^{-4\alpha_1/b} |1-u_a|^{-4\alpha_2/b} \Big) \, 
\prod_{1\leq a<a'}^{m}|u_a-u_{a'}|^{-4/b^{2}} =  \nonumber \\  
&& \ \ \,\Big( \pi \gamma(b^{-2})b^{2b^{-2}-2}\Big)^{b^2+1-b\sum_{i=1}^3\alpha_i} \frac{{\Upsilon_b'(0)\, }}{\Upsilon_b(\sum_{i=1}^3\alpha_i-Q)}
\prod_{i=1}^{3}\frac{\Upsilon_b(2\alpha_i)}{\Upsilon_b(\sum_{j=1}^3\alpha_j-2\alpha_i)} \label{FormulaA}
\end{eqnarray}
where $m/b+\alpha_1+\alpha_2+\alpha_3=Q$ with $Q=b+1/b$, and where the function $\Upsilon_b(x)$ is defined as \cite{ZZ}
\begin{equation}
    \log \Upsilon_b(x) = \int_{\mathbb{R}_{>0}} \, \frac{d\tau }{\tau }\, \Big[ \Big( \frac Q2 -x\Big)^2 e^{-\tau} - \frac{\sinh^2(\frac Q2 -x)\frac{\tau }{2}}{\sinh(\frac{b\tau}{2}) \sinh(\frac{\tau }{2b}) }\, \Big]\, .\label{HUPA}
\end{equation}
In (\ref{FormulaA}), $\Upsilon_b'(0)=\frac{d}{dx}\Upsilon_{b}(x)_{|x=0}$. Function $\Upsilon_b$ satisfies the reflection properties 
\begin{equation}
 \Upsilon_b(x) = \Upsilon_b(Q-x)\ \ , \ \ \ \ \  \Upsilon_b(x) = \Upsilon_{1/b}(x)
\end{equation}
and the shift properties
\begin{equation}
\Upsilon_b(x+b) = \gamma (xb)\, b^{1-2xb}\, \Upsilon_b(x) \ \ , \ \ \ \ \
\Upsilon_b(x+b^{-1}) = \gamma (xb^{-1})\, b^{-1+2xb^{-1}}\Upsilon_b(x)\, .\label{Shifta}
\end{equation}
Function $\Upsilon_b(x)$ has its zeroes at $x=mb+n/b$ with $m,n\in \mathbb{Z}_{>0}$ or $m,n\in \mathbb{Z}_{\leq 0}$.

As Eq. (\ref{dis}) states it, Liouville theory structure constants, $C(\alpha_1, \alpha_1, \alpha_3)$, are equal to Eq. (\ref{FormulaA}) above. Of course, the identity (\ref{FormulaA}) only makes sense for $m\in \mathbb{Z}_{>0}$, as in the l.h.s. of it $m$ is the number of integrals to be performed. However, as it is usual in this kind of CFT computation, we can first assume $m\in \mathbb{Z}_{>0}$ and then, after solving the Dotsenko-Fateev integral, analytically continue for values $\alpha_i\in\frac Q2 + i\mathbb{R}$ with $b\in\mathbb{R}$. This trick has shown to reproduce the correct result in many working examples, including timelike models with $b\in\mathbb{C}$ \cite{Yo4}, for which the analytic extension is much more subtle \cite{Witten2, Schom}. Using this method, formula (\ref{FormulaA}) can be easily obtained by iterating the shift equations (\ref{Shifta}) to first obtain
\begin{equation}
\prod_{r=1}^{n}\gamma (rb^2) = \frac{\Upsilon_b(nb+b)}{\Upsilon_b(b)}\, b^{n((n+1)b^2-1)} \ , \ \ \ \text{for}\, n \in \mathbb{Z}_{>0} \, ,
\end{equation}
and
\begin{equation}
\prod_{r=0}^{-n-1}\gamma (1+rb^2) = \frac{\Upsilon_b(nb+b)}{\Upsilon_b(b)}\, b^{n((n+1)b^2-1)} \ , \ \ \text{for}\, n \in \mathbb{Z}_{<0} \, .
\end{equation}
Using these and the other properties of $\Upsilon_b$, together with properties such as $\gamma(x)=1/\gamma(1-x)$ and $\gamma(x)\gamma(-x)=-1/x^2$, one finally arrives to (\ref{FormulaA}).

\subsubsection{Coulomb integral duality}

Another remarkable property of Dotsenko-Fateev integral that will be useful for us is the following recursion formula \cite{BF, FL, FL2}
\begin{eqnarray} 
&&\int_{\mathbb{C}^{n}}\prod_{i=1}^{n}d^2v_i \prod_{1\leq i <i'}^{n} |v_i - v_{i'}|^2 \prod_{i=1}^{n}\prod_{j=1}^{n+q+1} |v_i - x_j|^{2L_{j}} = \pi^{n-q}\, \frac{\Gamma(n+1)}{\Gamma(q+1)}\frac{\Gamma(-n-\sum_{j=1}^{n+q+1}L_j)}{\Gamma(1+n+\sum_{j=1}^{n+q+1}L_j)}   \nonumber \\
&&\, \prod_{j=1}^{n+q+1} \frac{\Gamma(1+L_j)}{\Gamma(-L_j)} \prod_{1\leq j <j'}^{n+q+1}|x_j - x_{j'}|^{2+2L_j+2L_{j'}} \int_{\mathbb{C}^2}\prod_{l=1}^{q}d^2v_l \prod_{1\leq l<l'}^{q} |v_{l}-v_{l'}|^2  \prod_{l=1}^{q} \prod_{j=1}^{n+q+1} |v_l - x_j|^{-2-2L_j} . \nonumber \\ \label{La}
\end{eqnarray}
This formula, which realizes the ``Coulomb integral duality'' mentioned in Figure 1, is exactly what will allow us to write the multiple integrals that appear in the sine-Liouville theory computation as standard conformal integrals of the type studied in the previous subsection. 

It was noticed by Fateev \cite{FateevUnpublished} that, by means of the integral duality formula (\ref{La}), one can write the sine-Liouville correlators (\ref{Integral}) as integrated correlators in standard Liouville theory. The precise relation follows from considering in (\ref{La}) the case $q=N-2-\Delta \omega $ and setting $z_N=\infty $. For instance, if in (\ref{Integral}) we consider $m_l=-\bar{m}_l$ for simplicity, then we find
\begin{eqnarray}
\mathcal{A}_{\text{sine-L}}^{N,\Delta \omega } &=& \frac{\lambda^{\Delta \omega }  \pi^{2+2\Delta \omega -N} }{b\Gamma (N-1-\Delta \omega )} \prod_{l=1}^N \frac{\Gamma(1+j_l-m_l)}{\Gamma (m_l-j_l)} \,  \prod_{1\leq l<l'}^{N-1}|z_l-z_{l'}|^{\frac 4k (m_l-\frac k2 ) (m_{l'}-\frac k2 )}  \nonumber \\
&&\, \prod_{1\leq l<l'}|z_l-z_{l'}|^{-4\alpha_l\alpha_{l'}}\, \int_{\mathbb{C}^q} \prod_{i=1}^{q}d^2v_i \, \prod_{1\leq i<i'}^{q} |v_i-v_{i'}|^{k} \prod_{i=1}^{q} \prod_{l=1}^{N-1}|z_l-v_i|^{2m_l-k} \,   \nonumber \\
&& \Big( {\pi b^4\lambda^2/\gamma(b^{-2})}\Big)^m \, b \, \Gamma(-m)\, \int_{\mathbb{C}^m}\prod_{a=1}^m d^2u_a \, \prod_{1\leq a<a'}^{m}|u_a-u_{a'}|^{-4b^{-2}}  \nonumber \\
&&\prod_{l=1}^{N-1}\prod_{a=1}^{m}|u_a-z_l|^{-4\alpha_l b^{-1}} \, \prod_{i=1}^q\prod_{a=1}^m |u_a-v_i|^{2b^{-2}}\label{FTF}
\end{eqnarray}
with $\alpha_l=b(k/2-1-j_l)$ for $l=1,2,...N$, $b^2=1/(k-2)$, with $m=-b\sum_{l=1}\alpha_l +b^2+(N-\Delta \omega )/2$, and $q=N-1+m=N-2-\Delta \omega $. The third and the fourth line of (\ref{FTF}) actually correspond to the Coulomb gas realization of a Liouville correlation function \cite{GL} involving $N$ exponential primary operators $V_{\alpha_l}=\exp (\sqrt{2}\alpha_l\varphi(z_l))$ and $N-2+\Delta \omega$ degenerate non-normalizable operators $V_{-\frac {1}{2b}}(v_i)$. The cosmological constant is given in terms of $\lambda$ by the relation $\pi b^2\lambda = (\pi\mu \gamma(b^2))^{1/(2b^2)}$. Coulomb gas computation (\ref{FTF}) involves the dual screening charge $V_{\frac 1b}(u)=\exp (\sqrt{2}b^{-1}\varphi (u))$, which implicitly invoke the Liouville self-duality under $b\leftrightarrow 1/b$, cf \cite{ZZ}. This means that we can write \cite{FateevUnpublished}
\begin{eqnarray}
\mathcal{A}_{\text{sine-L}}^{N,\Delta \omega } &=& \frac{\lambda^{\Delta \omega }  \pi^{2+2\Delta \omega -N} }{b\Gamma (N-1-\Delta \omega )} \prod_{l=1}^N \frac{\Gamma(1+j_l-m_l)}{\Gamma (m_l-j_l)} \,  \prod_{1\leq l<l'}^{N-1}|z_l-z_{l'}|^{\frac 4k (m_l-\frac k2 ) (m_{l'}-\frac k2 )}  \nonumber \\
&& \int_{\mathbb{C}^q}\, \prod_{i=1}^{N-2-\Delta \omega}d^2v_i \, \prod_{1\leq i<i'}^{N-2-\Delta \omega} |v_i-v_{i'}|^{k} \prod_{i=1}^{q} \prod_{l=1}^{N-1}|z_l-v_i|^{2m_l-k} \,  \, \nonumber \\
&&\, \times \, \, \Big\langle \ V_{\alpha_1}(z_1)\, \, ... \, \, V_{\alpha_{N-1}}(z_{N-1}) \, V_{\alpha_N}(\infty) \prod_{i=1}^{N-2-\Delta \omega} V_{-\frac{1}{2b}}(v_i)  \ \Big\rangle_{\text{L}}\, \label{FateevUn}
\end{eqnarray}
where the subscript $\text{L}$ on the r.h.s. stands for ``Liouville''. This formula, once FZZ correspondence is considered, turns out to be in agreement with the generalized $H_{3}^+$ WZW-Liouville correspondence; see Eq. (3.29) of \cite{Ribault}; see also Figure 3. 
\begin{figure}[h]
    \begin{center}
        \includegraphics[width=13cm]{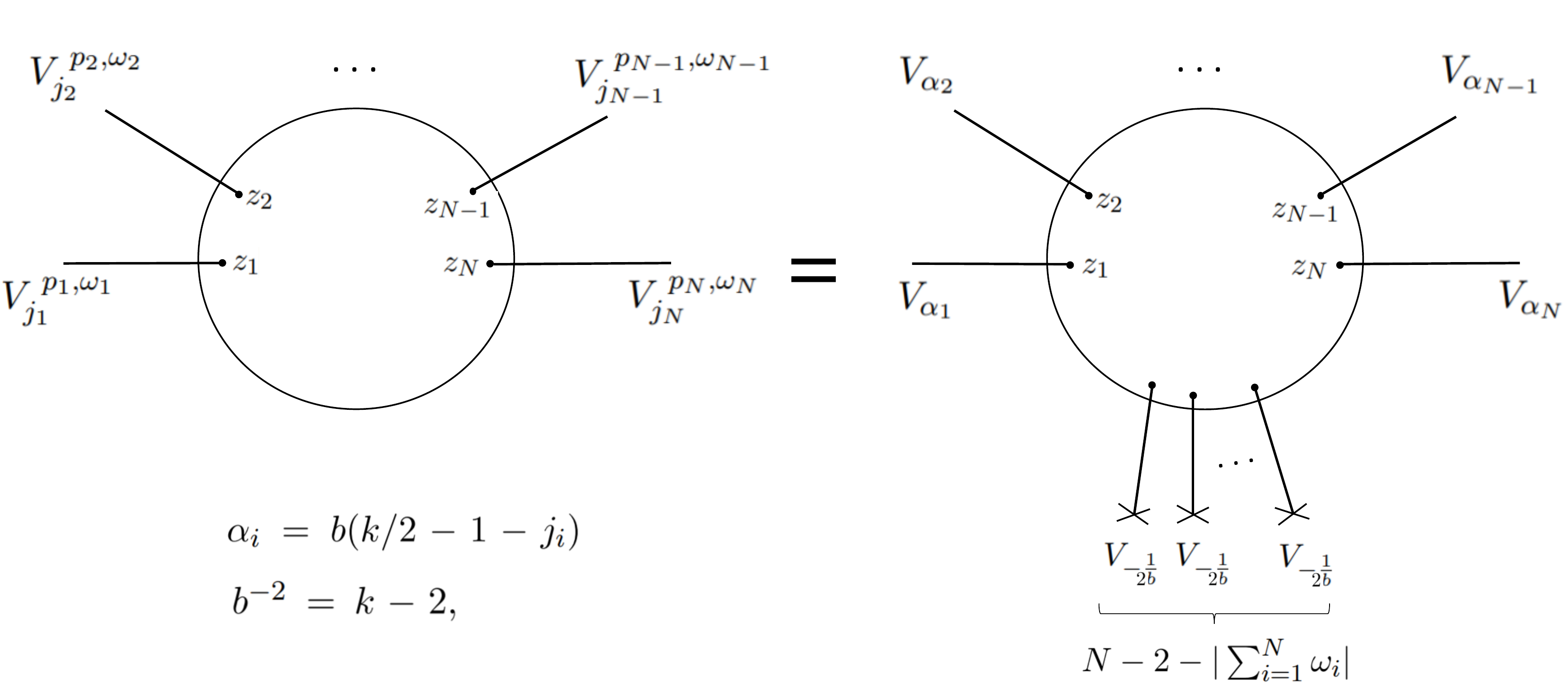}
        \caption{Scheme of the relation between the $N$-point functions in sine-Liouville theory and the ($2N-2-|\Delta \omega|$)-point function in Liouville field theory, where $N-2-|\Delta \omega|$ degenerate operators $V_{-\frac {1}{2b}}$ are integrated, together with an appropriate factor. Expression (\ref{FateevUn}) gives the precise relation.}
        \label{Ricci}
    \end{center}
\end{figure}

Here, we have considered $\Delta \omega \geq 0$. However, it is worth noticing that an analogous formula holds for the case $\Delta \omega \leq 0$ by simply replacing $m\leftrightarrow n$ and $u_a\leftrightarrow v_i$ in the procedure described above. In that case, the $\Gamma $ function in the denominator of the first factor in (\ref{FateevUn}) becomes $\Gamma (N-1-|\Delta \omega |)$. This implies that something special happens with the correlators that satisfy $|\Delta \omega | \geq N-2$. In that case, the mentioned $\Gamma $ function in the denominator diverges and the correlator vanishes. This is related to the bound on the violation of the total winding number in an $N$-point function; see, for instance, Appendix D of \cite{MO3}. 

\subsubsection{Sine-DOZZ redux}

We will apply the formulae above to study sine-Liouville 3- and 4-point functions. Let us start with the case $\Delta \omega = \sum_{i=1}^3\omega _i=1$; that is, $n=m+1$. For simplicity, let us assume $p_i=m_i+\bar m _i=0$ for $i=1,2,3$; namely, we have
\begin{equation}
\mathcal{A}^{N=3,\Delta \omega = 1}_{\text{sine-L}}\, =\, \Big\langle   \  V_{j_1, m_2, -m_1} (0) \, V_{j_2, m_2, -m_1} (1) \, V_{j_3, \frac k2-m_1-m_2, -\frac k2 +m_1+m_2} (\infty) \ \Big\rangle_{\text{sine-L}}
\end{equation}
In this case, we get
\begin{eqnarray}
\mathcal{A}^{N=3,\Delta \omega = 1}_{\text{sine-L}} &=& \lambda^{2m+1}\, \frac{\Gamma (-2m-1) \Gamma (2m+2)}{\Gamma (m+2)\Gamma(m+1)}\int_{\mathbb{C}^{m}} \prod_{a=1}^{m}d^2u_a \int_{\mathbb{C}^{n}} \prod_{i=1}^{n}d^2v_i \, 
\prod_{a=1}^{m} \prod_{i=1}^{n}  |u_a-v_{i}|^{2-2k}
 \nonumber \\
&&\ \ 
\prod_{a=1}^{m} \Big[ |u_a|^{2(j_1+m_1)}  |1-u_a|^{2(j_2+m_2)}\Big]\,
\prod_{i=1}^{n} \Big[ |v_i|^{2(j_1-m_1)}|1-v_i|^{2(j_2-m_2)}
\Big]\,
\nonumber \\
&& \ \ \prod_{1\leq a<a'}^{m}  |u_a-u_{a'}|^{2} \, \prod_{1\leq i<i'}^{n}  |v_i-v_{i'}|^{2}  \, ,
\end{eqnarray}
where $m$ is given by the equation $j_1+j_2+j_3+1=(k-2)(m+\frac 12)$.

Considering formula (\ref{La}) to integrate over the variables $v_i$, which corresponds to the case $q=0$; $x_1=0$, $x_2=1$, $x_{3}=u_1$, ... $x_{n+1}=u_{n-1}$ ; $L_1 = j_1-m_1$, $L_2=j_2-m_2$, $L_3=L_4=...=L_{n+1}=1-k$; we obtain 
\begin{eqnarray}
\mathcal{A}^{N=3,\Delta\omega = 1}_{\text{sine-L}} &=& \lambda^{2m+1}\, \frac{\Gamma (-2m-1) \Gamma (2m+2)}{\Gamma (m+1)\Gamma(-m)} \pi^{m+1}\, \Big( \frac{\Gamma(2-k)}{\Gamma(k-1)} \Big)^{m}\nonumber \\
&& \Gamma(-m)\int_{\mathbb{C}^{m}} \prod_{a=1}^{m}d^2u_a \, \prod_{a=1}^{m} \Big[ |u_a|^{4(j_1+1- k/2 )}  |1-u_a|^{4(j_2+1-k/2 )} \Big] \, \prod_{1\leq a<a'}^{m}  |u_a-u_{a'}|^{4(2-k)}  \, .\nonumber
\end{eqnarray}
The second line of the r.h.s. of this equation corresponds to a 3-point function of Liouville field theory. As before, the Liouville central charge is $c=1+6Q^2>25$ with $Q=b+1/b$ and $b^{-2}=k-2$, and the Liouville momenta are $\alpha_i=b(k/2 -1-j_i)$ for $i=1,2,3$. The number of integrals is $m=-b\sum_{i=1}^3\alpha_i+b^2+1$; these integrals represent the insertion of $m$ screening operators $V_{\frac 1b}(u_a)=\exp (\sqrt{2}b^{-1}\varphi (u_a))$. This yields
\begin{equation}
\mathcal{A}^{N=3,\Delta\omega = 1}_{\text{sine-L}} = \frac{\pi \lambda }{b} \, \prod_{i=1}^{3}\frac{\Gamma (1+j_i-m_i) }{\Gamma(m_i-j_i)}\, \,  \Big\langle V_{\alpha_1}(0)V_{\alpha_2}(1)V_{\alpha_3}(\infty)\Big\rangle_{\text{L}} \label{JKL77}
\end{equation}
which directly relates the winding violating sine-Liouville 3-point function with the Liouville DOZZ structure constant. The KPZ scaling is, therefore, $\mathcal{A}^{N=3,\Delta\omega = 1}_{\text{sine-L}}\sim \lambda ^{\frac{2}{k-2}(\sum_{i=1}^3j_i+1)}\sim \lambda \mu^{(Q-\sum_{i=1}^3\alpha_i)/b}$.

Using the formulae above and recovering all the factors, the final result reads
\begin{eqnarray}
\mathcal{A}^{N=3,\Delta\omega = 1}_{\text{sine-L}} &=& {\Big(  \lambda \pi b^{2+b^{-2}} \Big) }^{\frac{2}{k-2}(\sum_{i=1}^3j_i+1)}\, 
\frac{\Gamma (\frac 12 -b^2 (\sum_{i=1}^3j_i +1)) }{\Gamma (\frac 12 +b^2 (\sum_{i=1}^3j_i +1)) } \prod_{t=1}^{3}
\frac{\Gamma (1+j_t-m_t)}{\Gamma (m_t-j_t)}
\nonumber \\ 
&&\ \ \ \  \ \ \frac{b^{-2-b^{-2}} \pi^{-2}\,  \Upsilon_b'(0)}{ \Upsilon_b(\frac{1}{2b} + b(\sum_{i=1}^{3} j_i +1))} \prod_{i=1}^{3} \frac{\Upsilon_b(b(2j_i+1))}{\Upsilon_b (\frac{1}{2b}-b(\sum_{l=1}^{3} j_l-2j_i))}  \, \, 
 \, ,
\end{eqnarray}
where we have used $\Gamma(-2m-1)\Gamma(2m+2)=(-1)^{m}\Gamma(m+1)\Gamma(-m) $ and properties of $\gamma $.

In comparison with the result of \cite{FH}, here we find an additional factor $\frac{\Gamma(\frac 12 -b^2(\sum_{i=1}^3j_i+1))}{\Gamma(\frac 12 +b^2(\sum_{i=1}^3j_i+1))}$, which exhibits poles at $\sum_{i=1}^{3}j_i+1=(k-2)(m+\frac 12)$ with $m\in \mathbb{Z}_{\geq 0}$. These poles correspond to resonant correlators, in which the number of screening operators in a non-negative integer, cf. \cite{diFK}. This is well explained if we look at the integration over the zero mode of the Liouville type field, which produces an infinite factor due to the non-compactness of the target space. To see this explicitly, we can consider the limit $\lim _{\varepsilon\to 0}\frac{\Gamma(1+m+\varepsilon )\Gamma(-m-\varepsilon)}{\Gamma(-\varepsilon)}= (-1)^{m}$, which transforms the overall factor $\frac{1}{m!}$ appearing in the residue of the correlation functions into the factor $\Gamma(-m)$ of the actual correlation functions when one takes into account the infinite factor that comes from the integration over the zero mode; see Eqs. (2.10)-(2.12) of \cite{diFK}, Eq. (3.3) of \cite{GL}, and Eq. (2.23) of \cite{KKK} for details.

\subsubsection{$H_3^+$ WZW-Liouville from FZZ}

Now, let us consider the case $\Delta \omega = \sum_{i=1}^3\omega_i=0$. Repeating the procedure above, we can also find a closed expression for this observable in terms of Liouville correlators: Considering the 3-point funciton of the sine-Liouville field theory for $m=n$, and using the formula (\ref{La}) to integrate over the variables $v_i$, now with $q=1$, one finds that the result is
\begin{eqnarray}
\mathcal{A}^{N=3,\Delta\omega = 0} &=& \frac{1}{\pi} \Big( {\pi b^4\lambda^2 }{\gamma(1-b^{-2})}\Big)^n\, \prod_{i=1}^{3}\gamma (1+j_i-m_i) \Gamma(-n)  \int_{\mathbb{C}} d^2v  |v|^{2(m_1-j_1-1)} |1-v|^{2(m_2-j_2-1)}  
\nonumber \\
&& \ \int_{\mathbb{C}^{n}} \prod_{a=1}^n\, d^2u_a\, \prod_{1\leq a<a'}^{n} |u_a-u_{a'}|^{-4/b^{2}}\, \prod_{a=1}^{n}|u_a-v|^{2/b^{2}} \prod_{a=1}^{n} \Big[ |u_a|^{-4\alpha_1 /b}|1-u_a|^{-2\alpha_2 /b} \Big] \,  \nonumber
\end{eqnarray}
where, now, we have $n/b+\sum_{i=1}^3\alpha_i-{1}/({2b})=Q$. This can be seen to correspond to and integrated 4-point function of Liouville theory with the insertion of a degenerate operator $V_{-\frac{1}{2b}}(v)$. More precisely, it reads
\begin{equation}
\mathcal{A}^{N=3,\Delta\omega = 0} = \frac{1}{\pi b} \prod_{i=1}^{3}\frac{\Gamma (1+j_i-m_i)}{\Gamma(m_i-j_i)} \, \int_{\mathbb{C}} d^2v \, |v|^{2m_1-k}|1-v|^{2m_2-k} \, \Big\langle \ V_{\alpha_1}(0)\, V_{\alpha_2}(1)\, V_{-\frac{1}{2b}}(v)V_{\alpha_3}(\infty)\ \Big\rangle_{\text{L}} \nonumber
\end{equation}
where, as before, $\alpha_i=b(k/2-1-j_i)$, $i=1,2,3$, and $b^2=1/(k-2)$. It is worth noticing that the KPZ scaling is the correct one: The Liouville cosmological constant scales as $\mu \sim {\lambda}^{2b^2}$, so we find $\mathcal{A}^{N=3,\Delta\omega = 0} \sim \mu^{(Q-\sum_{i=1}^4\alpha_i)/b}$ with the fourth momentum being $\alpha_4=-1/2b$; cf. Eq. (3.2)-(3.4) in \cite{ZZ}. The same scaling is found in all winding preserving correlators, i.e. those $N$-point correlators with $\Delta \omega =0$, for which $\sum_{i=1}^Nj_i=-\sum_{i=1}^N\alpha_i/b+N/(2b^2)$. In the cases in which the winding number is violated in $\Delta \omega $ units, there is an extra factor $\mu^{\Delta \omega/(2b^2)}\sim \lambda^{\Delta \omega}$ on the r.h.s.. The precise factor is given by (\ref{FateevUn}), and, in particular, it explains the factor $c_b^r$ appearing in the expressions of \cite{Ribault} and some $k$-dependent factors omitted in \cite{MO3} (to compare with the result of \cite{MO3} one has to take into account the definition of the special function $G(x)=\Upsilon^{-1}_b(-xb)b^{-x^2b^2-x(b^2+1)}$, which can also de defined in terms of double gamma functions, $\Gamma_2$, or Barnes $G$-functions).

The expression above agrees with the $H_3^+$ WZW-Liouville correspondence \cite{RT, Ribault, HS}. To compare with \cite{Ribault} it is necessary to consider the Weyl-reflected convention $j\to -1-j$; to compare with \cite{RT} it is necessary, in addition, to change $m\to -m$ and $\bar m \to -\bar m$; as we already noticed, all these are symmetries of the formulae of the conformal dimensions $h,\bar h $, so it is matter of convention.

\section{The 4-point function}

\subsubsection{Maximally winding violating correlators}

As said, we can use similar techniques to consider the case with arbitrary number of insertions. For example, consider the maximally winding violating 4-point function
\begin{equation}
\mathcal{A}^{N=4,\Delta\omega = 2}_{\text{sine-L}} \, = \, \Big\langle \ V_{j_1,m_1, \bar{m}_1}(0)\, V_{j_2,m_2, \bar{m}_2}(1)\, V_{j_3,m_3, \bar{m}_3}(z)\, V_{j_4,k -m_1-m_2-m_3, -k + {m}_1+ {m}_2+ {m}_3} (\infty)\ \Big\rangle_{\text{sine-L}}\nonumber
\end{equation}
Proceeding as before, but now for the case $N=4$ with $\Delta \omega = \sum_{i=1}^4\omega_i=2$, we find
\begin{eqnarray}
\mathcal{A}^{N=4,\Delta\omega = 2}_{\text{sine-L}} &=& \pi^2\lambda^2 \, \Big( {\pi b^4\lambda^2 }{\gamma(1-b^{-2})}\Big)^m\, \Gamma(-m)\, \prod_{i=1}^4\frac{\Gamma (1+j_i-m_i)}{\Gamma({m}_i-j_i)}|z|^{\frac 4k (m_1-\frac k2)(m_3-\frac k2)-4\alpha_1\alpha_3}
 \,  \nonumber \\
&& \, |1-z|^{\frac 4k (m_2-\frac k2)(m_3-\frac k2)-4\alpha_2\alpha_3} \int_{\mathbb{C}^{m}} \prod_{a=1}^{m}d^2u_a  \prod_{1\leq a<a'}^{m}  |u_a-u_{a'}|^{-4/b^{2}}\nonumber \\
&&\, \prod_{a=1}^{m} \Big[ |u_a|^{-4\alpha_1 /b}  |1-u_a|^{-4\alpha_2/ b} |z-u_a|^{-4\alpha_3/ b} \Big] \, \label{OPOOO}
\end{eqnarray}
where, now, $m/b=-\sum_{i=1}^4\alpha_i +b+1/b$. That is,
\begin{eqnarray}
\mathcal{A}^{N=4,\Delta\omega = 2}_{\text{sine-L}} \, &=& \, \frac{\pi^2\lambda ^2}{b} \, |z|^{\frac 4k (m_1-\frac k2)(m_3-\frac k2)} |1-z|^{\frac 4k (m_2-\frac k2)(m_3-\frac k2)} \,  \prod_{i=1}^4\frac{\Gamma (1+j_i-m_i)}{\Gamma(m_i-j_i)}\,\nonumber \\
&&\, \ \ \,\times \, \, \Big\langle \ V_{\alpha_1}(0)\, V_{\alpha_2}(1)\, V_{\alpha_3}(z)\, V_{\alpha_4}(\infty) \ \Big\rangle_{\text{L}} \label{El38estacargado}
\end{eqnarray}

This means that, as it happens with the 3-point function, the maximally violating sine-Liouville 4-point function $\mathcal{A}^{N=4,\Delta\omega = 2}_{\text{sine-L}} $ turns out to be proportional to a Liouville 4-point function. Once again, this is in agreement with the expectation coming from the generalization of the $H_3^+$ WZW-Liouville correspondence to the case in which the spectral flow symmetry of $SL(2,\mathbb{R})_k$ is taken into account; see Eq. (3.29) of \cite{Ribault}. 

In particular, formula (\ref{El38estacargado}) allows us to study the crossing symmetry and factorization properties of the maximally winding violating 4-point function in sine-Liouville theory by means of using the properties of Liouville theory conformal blocks. For instance, it permits to observe that the dependence on the cross-ratio $z$ around $z\simeq 0$ is
\begin{eqnarray}
\mathcal{A}^{N=4,\Delta\omega = 2}_{\text{sine-L}} \, \sim \, |z|^{2\delta} \,  
\int_{\mathbb{C}^{m}} \prod_{a=1}^{m}d^2w_a  \prod_{1\leq a<a'}^{m}  |w_a-w_{a'}|^{-4/b^{2}}\, \prod_{a=1}^{m} \Big[ |w_a|^{-4\alpha_1 /b} |1-w_a|^{-4\alpha_3/ b}  |1-z\,w_a|^{-4\alpha_2/ b} \Big] \,\nonumber
\end{eqnarray}
with $\delta = \frac 2k (m_1-\frac k2)(m_3-\frac k2)+(\Delta_2+\Delta_4-\Delta_1-\Delta_3)$, where $\Delta_i = \alpha_i (Q-\alpha_i)=-\frac{j_i(j_i+1)}{k-2}+\frac k4$ are the conformal dimension of the Liouville primaries $V_{\alpha_i}(z_i)$. More interestingly, we can write the following formula for the sine-Liouville conformal blocks in the case $\Delta \omega=2$,
\begin{eqnarray}
\mathcal{A}^{N=4,\Delta\omega = 2}_{\text{sine-L}} & = & \, \frac{\pi^2\lambda ^2}{b} \, \,
\prod_{i=1}^4\frac{\Gamma (1+j_i-m_i)}{\Gamma({m}_i-j_i)}\, \, |z|^{2(\Delta _2 -\Delta _3 - \Delta _1)} \, \, |1-z|^{2(\Delta _1 -\Delta _2 - \Delta _3)}
\,
\nonumber \\
&&  \int_{j\in -\frac 12 +i\mathbb{R}}dj \ \ C\Big( \frac k2 -1-j_1, \frac k2 -1-j_2, \frac k2 +j\Big)\, \Big| \mathcal{F}\Big( \frac k2 -1-j_i ,  \frac k2 -1 -j , z\Big)\Big| ^2\, \nonumber \\
&& \ \ \ \  \ \ \ \ \ \ \ \ \ \ \ \ \,  \,   \, C\Big( \frac k2 -1-j,  \frac k2 -1-j_3, \frac k2 -1-j_4\Big)\,   
 \, \label{J36}
\end{eqnarray}
where $\mathcal{F}(\alpha_i , \alpha , z)$ and $C(\alpha_1, \alpha_2, \alpha_3 )$ are the conformal blocks and the structure constants of Liouville field theory, respectively; see Eqs. (2.23) and (3.14) of \cite{ZZ}. Notice that, in this expression, the integral on normalizable intermediate states $j\in -\frac 12 +i\mathbb{R}$ emerges naturally from the integral over the intermediate normalizable Liouville momenta $\alpha = b(k/2-1-j)\in \frac Q2 +i\mathbb{R}$ in the Liouville 4-point function. This actually proves the crossing symmetry of the maximally violating amplitude.

Of course, one could argue that expression (\ref{J36}) was already implicitly encoded in the $H_3^+$ WZW-Liouville correspondence. However, let us be reminded of the fact that the maximally winding violating amplitudes --i.e. those with $\Delta \omega =N-2$, like in (\ref{J36})-- remained in \cite{Ribault} as a conjecture due to the absence of degenerate fields $V_{-\frac{1}{2b}}$ in the corresponding Liouville correlator, what made impossible to get information from the Belavin-Polyakov-Zamolodchikov decoupling equation. Here, we are deriving (\ref{J36}) directly from the sine-Liouville theory Coulomb gas computation. Strictly speaking, the fact that this sine-Liouville computation yields a result in agreement with the conjecture of \cite{Ribault} for the case $\Delta \omega =N-2$ can be considered as a consistency check of the latter. 

\subsubsection{Winding preserving correlators}

As a further consistency check of the whole procedure, we can follow the same procedure to compute the winding preserving sine-Liouville 4-point function. This yields
\begin{eqnarray}
\mathcal{A}^{N=4,\Delta\omega = 0}_{\text{sine-L}} &= &\frac 12 \Big( {\pi b^4\lambda^2 }{\gamma(1-b^{-2})}\Big)^n\,  
\prod_{i=1}^{4}\frac{\Gamma (1+j_i-m_i)}{\Gamma({m}_i-j_i)} \, \, \Gamma(-n) \, |z|^{\frac 4k (m_1-\frac k2 )(m_3-\frac k2 )-4\alpha_1\alpha_3}  \nonumber \\
&& \,
\, |1-z|^{\frac 4k (m_2-\frac k2 )(m_3-\frac k2 )-4\alpha_2\alpha_3} \, \int_{\mathbb{C}} d^2v_1\, \int_{\mathbb{C}} d^2v_2 \, \, |v_1-v_2|^{2}\,\prod_{i=1}^2\Big[ |v_i|^{2m_1-k} 
\nonumber \\
&& \,|1-v_i|^{2m_2-k} |z-v_i|^{2m_3-k} \Big]  \, \int_{\mathbb{C}^{n}} \prod_{a=1}^n\, d^2u_a\, \prod_{1\leq a<a'}^{n} |u_a-u_{a'}|^{-4/b^{2}}\,   
\nonumber \\
&& 
\prod_{a=1}^{n}\prod_{i=1}^2|u_a-v_i|^{2/b^{2}}\prod_{a=1}^{n} \Big[\, |u_a|^{-4\alpha_1 /b}|1-u_a|^{-2\alpha_2 / b} |z-u_a|^{-2\alpha_3 /b}\, \Big] \, ; \label{ZetaYu}
\end{eqnarray}
where $n=b^2(\sum_{i=1}^4j_i+1)= -b\sum_{i=1}^4\alpha_i+b^2+2$. This yields
\begin{eqnarray}
\mathcal{A}^{N=4,\Delta\omega = 0}_{\text{sine-L}} &=& \frac{1}{2b\pi^2} \prod_{i=1}^{4}\frac{\Gamma (1+j_i-m_i)}{\Gamma({m}_i-j_i)}
\, |z|^{\frac 4k (m_1-\frac k2 )(m_3-\frac k2 )} 
\, |1-z|^{\frac 4k (m_2-\frac k2 )(m_3-\frac k2 )}\nonumber \\
&&\, \int_{\mathbb{C}} d^2v_1 \int_{\mathbb{C}} d^2v_2 \, \, |v_1-v_2|^{k}\, \prod_{i=1}^2\Big[ |v_i|^{2m_1-k}|1-v_i|^{2m_2-k} |z-v_i|^{2m_3-k} \Big] \, \nonumber \\
&&
\, \,\times \, \, \Big\langle \ V_{\alpha_1}(0)\, V_{\alpha_2}(1)\, V_{\alpha_3}(z)\, V_{\alpha_4}(\infty)\, V_{-\frac{1}{2b}}(v_1)\, V_{-\frac{1}{2b}}(v_2)\, \ \Big\rangle_{\text{L}}  \, 
\end{eqnarray}
which shows that the sine-Liouville 4-point function $\mathcal{A}^{N=4,\Delta\omega = 0}_{\text{sine-L}}$ is proportional to an integrated Liouville 6-point, with four operators of momenta $\alpha_i=b(k/2-j_i-1)$ ($i=1,2,3,4$) and two degenerate operators of momentum $\alpha_{5,6} = -1/(2b)$, with $b^2=1/(k-2)$. Again, this is in complete agreement with the $H_3^+$ WZW-Liouville correspondence, cf. \cite{Ribault}. In fact, the proof of the such correspondence given in \cite{HS} also resorts to the Liouville self duality under $b \leftrightarrow b^{-1}$ like the computation we discussed above, the difference being that the computation in  \cite{HS} is done in the path integral formalism, and so it achieves to make it for arbitrary genus, while the one presented here is on the sphere topology but is also valid for arbitrary winding number. In this sense, both computations are complementary.

\subsubsection{Other processes}

The last consistency check for the 4-point function would be to work out the intermediate case between (\ref{OPOOO}) and (\ref{ZetaY}); that is, the case $\Delta \omega = 1$. With the same procedure, we find
\begin{eqnarray}
\mathcal{A}^{N=4,\Delta\omega = 1}_{\text{sine-L}} &= &\lambda \Big( {\pi b^4\lambda^2 }{\gamma(1-b^{-2})}\Big)^n\,  
\prod_{i=1}^{4}\frac{\Gamma (1+j_i-m_i)}{\Gamma({m}_i-j_i)} \, \Gamma(-n) \,  \,
|z|^{\frac 4k (m_1-\frac k2 )(m_3-\frac k2 )-4\alpha_1\alpha_3 }\, \nonumber \\
&&|1-z|^{\frac 4k (m_2-\frac k2 )(m_3-\frac k2 )-4\alpha_2\alpha_3 } 
\, 
\int_{\mathbb{C}} d^2v \, \Big[ |v|^{2m_1-k}\, |1-v|^{2m_2-k} \, |z-v|^{2m_3-k} \Big]
\nonumber \\
&& \,   \, \int_{\mathbb{C}^{n}} \prod_{a=1}^n\, d^2u_a\, \prod_{1\leq a<a'}^{n} |u_a-u_{a'}|^{-4/b^{2}}\,   \prod_{a=1}^{n}|u_a-v|^{2/b^{2}} \, \, \prod_{a=1}^{n} \Big[\, |u_a|^{-4\alpha_1 /b}
\nonumber \\
&& 
\ \ \ \ \, |1-u_a|^{-2\alpha_2 / b} |z-u_a|^{-2\alpha_3 / b}\, \Big] \, 
, \label{ZetaY}
\end{eqnarray}
with $n=-b\sum_{i=1}\alpha_i +b^2+3/2$; which can be written as
\begin{eqnarray}
\mathcal{A}^{N=4,\Delta\omega = 1}_{\text{sine-L}} &= &  \frac{\lambda }{b} \, \prod_{i=1}^{4}\frac{\Gamma (1+j_i-m_i)}{\Gamma({m}_i-j_i)} \, 
|z|^{\frac 4k (m_1-\frac k2 )(m_3-\frac k2 )}\, |1-z|^{\frac 4k (m_2-\frac k2 )(m_3-\frac k2 )} 
\nonumber \\
&& \int_{\mathbb{C}} d^2v\, \Big[ |v|^{2m_1-k}|1-v|^{2m_2-k}\, |z-v|^{2m_3-k} \Big]  \,  
\nonumber \\
&&  \  \  \  \ \,\times \, \, \Big\langle \ V_{\alpha_1}(0)\, V_{\alpha_2}(1)\, V_{\alpha_3}(z)\, V_{\alpha_4}(\infty)\, V_{-\frac{1}{2b}}(v)\, \ \Big\rangle_{\text{L}} \, . \label{ZetaYdd}
\end{eqnarray}
Again, this is in complete agreement with the $H_3^+$ WZW-Liouville correspondence \cite{Ribault} once FZZ duality is considered. In the cases in which $\bar m _i\neq -m_i$, a similar formula holds; we just need to replace the $\Gamma$ functions in the denominator of (\ref{ZetaYdd}) by $\Gamma(\bar m_i-j_i)$ and split the factors as follows
\begin{eqnarray}
|v_i-z_l|^{2m_i-k} &\to&  (v_i-z_i)^{m_i-\frac k2}(\bar v _i-\bar z _i)^{\bar{m}_i-\frac k2}\nonumber \\
|z_l -z_{l'}|^{\frac 4k(m_l-\frac k2)(m_{l'}-\frac k2)}&\to& 
(z_l -z_{l'})^{\frac 2k(m_l-\frac k2)(m_{l'}-\frac k2)}
(\bar z _l -\bar z _{l'})^{\frac 2k(\bar m _l-\frac k2)(\bar m_{l'}-\frac k2)}\nonumber
\end{eqnarray}

The final expressions for the AdS$_3$ correlators follow from multiplying the sine-Liouville expressions above by the prefactor 
\begin{equation}
(z)^{\frac 2k (m_1+\frac k2 \omega_1)(m_3+\frac k2 \omega_3)} (1-z)^{\frac 2k (m_2+\frac k2 \omega_2)(m_3+\frac k2 \omega_3)}
(\bar{z})^{\frac 2k (\bar{m}_1+\frac k2 \omega_1)(\bar{m}_3+\frac k2 \omega_3)} (1-\bar z )^{\frac 2k (\bar{m}_2+\frac k2 \omega_2)(\bar{m}_3+\frac k2 \omega_3)}\nonumber 
\end{equation}

In conclusion, the sine-Liouville 4-point functions (\ref{ZetaYu}), (\ref{ZetaY}) and (\ref{OPOOO}) give the tree-level 4-point correlators on AdS$_3$ for the cases $\Delta \omega =0$, $\Delta \omega =1$ and $\Delta \omega =2$, respectively. 

\[\]

This work has been supported by CONICET and ANPCyT through grants PIP-1109-2017, PICT-2019-00303.

  \end{document}